\documentclass[12pt]{article}

\usepackage{latexsym,amsfonts,amsmath,amsthm,amssymb,amsbsy,multirow ,textcomp,hyperref,wrapfig,datetime,verbatim,cancel,subfigure,cite,comment}
\usepackage[dvipdfmx]{graphicx}
\usepackage[font={footnotesize,sl},bf]{caption}
\usepackage{epstopdf}
\usepackage{enumerate}
\usepackage{tensor}
\usepackage{slashed}
\usepackage{feynmf}
\usepackage{hyperref}
\usepackage{putex}

\usepackage[usenames,dvipsnames,svgnames,table]{xcolor}

\hypersetup{
	pdfstartview={FitH},    
	colorlinks=true,       
	linkcolor=blue,          
	citecolor=blue!59!black! ,        
	filecolor=magenta,      
	urlcolor=blue!75!black,           
	linktoc=all
}

\def\be{\begin{equation}}
\def\ee{\end{equation}}
\def\bea{\begin{eqnarray}}
\def\eea{\end{eqnarray}}
\newcommand{\beq}{\begin{eqnarray}}
\newcommand{\eeq}{\end{eqnarray}}

\newcommand{\ba}{\begin{align}}
\newcommand{\ea}{\end{align}}

\numberwithin{equation}{section}

\long\def\new#1\endnew{{\bf #1}}		
\long\def\del#1\enddel{}

\def\del{\partial}

\usepackage{color}
\usepackage[dvipsnames]{xcolor}

\newcommand{\pink}[1]{\textcolor{\pink}{#1}}

\def\tr{\text{tr}}

\def\Z{\mathbb{Z}}
\def\({\begin{equation}}
\def\){\end{equation}}

\newcommand{\ket}[1]{\left| #1\right>}
\newcommand{\bra}[1]{\left< #1\right|}

\begin{document}

	\institution{UCLA}{ \ \quad\quad\quad\quad\ \,  Mani L. Bhaumik Institute for Theoretical Physics
		\cr Department of Physics \& Astronomy,\,University of California,\,Los Angeles}

	\title{Universality in the OPE Coefficients of Holographic 2d CFTs
	}
	
	\authors{Ben Michel}
\date{}

\abstract{The thermodynamic stability of large AdS$_3$ black holes implies that Cardy's $\Delta\rightarrow\infty$ formula for the density of states remains approximately valid when $\Delta\sim c$ in holographic 2d CFTs, constraining their light spectra. Averaged OPE coefficients take a similarly universal asymptotic form, and black hole arguments again imply an extended regime of validity. In this note we study conditions under which the OPE asymptotics extend to $\Delta\sim c$ at large central charge. Some of the conditions found are stronger than required by an extended Cardy regime and are violated by permutation orbifolds, such as the D1-D5 system at zero coupling. Our results suggest new bounds on non-vacuum block contributions to correlation functions in holographic CFTs.}

	\maketitle
	\setcounter{tocdepth}{2}
	\begingroup
	\hypersetup{linkcolor=black}
	\tableofcontents
	\endgroup

\section{Introduction}
Holography implies many
phenomena that are non-generic in the space of conformal field theories. A prime example in two dimensions
is the extended regime of validity of Cardy's formula~\cite{Cardy:1986ie}: modular invariance
determines a universal form for the entropy as $\Delta\rightarrow\infty$, but the thermodynamic stability of large AdS$_3$ black holes implies that this form remains valid for all~$\Delta>c/6$ if the theory has a semiclassical bulk dual. On the field theory side this phenomenon was explained by Hartman, Keller and Stoica (HKS)~\cite{Hartman:2014oaa}: a more refined modular invariance argument shows that Cardy's formula extends to all such $\Delta$ in any
CFT with large $c$ and sufficiently sparse light spectrum.

However, the bulk does not just count states. Modular properties constrain other asymptotic CFT data to similarly universal forms, including the averaged OPE coefficients
$\overline{C_{ijk}^2}$ in the limit where at least one of the dimensions
$\Delta_i$, $\Delta_j$, $\Delta_k\rightarrow\infty$. This data involves black holes in the bulk, and holographic reasoning
may again suggest an extended regime of validity.
In this note we study the conditions under which these formulas (and others) extend to the regime where at least one of the operators has $\Delta>c/6$ by adapting the techniques of \cite{Hartman:2014oaa} to the
quantities that encode their asymptotics.

Denoting light and heavy operators by $L$ and $H$, the results of
\cite{Das:2017cnv}, \cite{Brehm:2018ipf} and \cite{Cardy:2017qhl} for the squared OPE
coefficients at equal operator dimensions are respectively
\beq
\label{hlleq}
\overline{C_{HLL}^2} &\approx&
16^{-\Delta_H}  \cdot e^{-S_\text{BH}(\Delta_H)/2} \\
\label{hhleq}
\overline{C_{HHL}^2} &\approx& e^{-S_\text{BH}(\Delta_H)}\\
\label{hhheq}
\overline{C_{HHH}^2} &\approx& e^{-3S_\text{BH}(\Delta_H)/2}
\eeq
as $\Delta_H\rightarrow \infty$, where $S_\text{BH}$ is the Cardy entropy
\(
{S_\text{BH}} = {2\pi\sqrt{\frac{c}{3}\left(\Delta-\frac{c}{12}\right)}}
\)
and the average is taken over all states with dimension
$\Delta_H$.\footnote{In \cite{Das:2017cnv} the average is taken over primaries rather
  than all states, and the result takes the form above with $c\rightarrow c-1$ as in the refined Cardy
  formula for the density of primary states \cite{Kraus:2016nwo}. In \cite{Cardy:2017qhl} the average is also taken over primaries but they work at large $c$, so the shift is invisible. We will unrefine
  these results to averages over all states below.} We will find that \eqref{hlleq} remains valid for $\Delta>c/6$ if
$\rho(\Delta)\lesssim e^{\pi\Delta}$ when $\Delta<c/12+\epsilon$, a condition that is violated
by all permutation orbifolds including the free D1-D5 CFT. This quantity is captured by some bulk process but a precise argument has
not been made, and it is unclear if this condition is
implied by holography. However there is a sharp bulk argument that
\eqref{hhleq} remains valid in the same regime, and we show that
under some mild additional assumptions
(stated in footnote~\ref{footnote}) the weaker HKS sparseness condition $\rho(\Delta)\lesssim
e^{2\pi\Delta}$ suffices. These conditions also
guarantee that asymptotic formulas for $\overline{C_{HHL}}$ and the
density of primary states remain valid in the extended regime. We discuss
conditions under which \eqref{hhheq} might extend as well in section~\ref{sec:chhh}.

The rest of this note is organized as follows. In section~\ref{generalities} we adapt the approach of HKS to modular covariant
quantities. In section~\ref{extended} we review the derivations of the asymptotic
expressions for $\overline{C_{HLL}^2}$, $\overline{C_{HHL}^2}$,
$\overline{C_{HHL}}$ and the density of primary states, and find
conditions under which their regimes of validity extend to all $\Delta_H>c/6$. In section~\ref{further} we
discuss obstacles to extending formulas for
$\overline{C_{HHH}^2}$ and averages of
$C_{ijk}^2$ over primaries. The pieces come together for a discussion
of vacuum block dominance, section~\ref{vbd}. Results for averages over all operators with fixed dimension and spin can be found in the appendix, where polynomial prefactors are also kept.

\section{Covariant HKS}
\label{generalities}

The HKS argument \cite{Hartman:2014oaa} uses modular invariance to show that Cardy's formula for the density of states as $\Delta\rightarrow\infty$ remains valid for all $\Delta >c/6$ in large $c$ CFTs with a sufficiently sparse light spectrum. The objects that encode the asymptotics of OPE coefficients are instead modular covariant, so a generalization of their argument will be needed.

Suppose we have a quantity $X(\beta)$ of the form

\(
\label{boltz}
X(\beta) = \sum_{\text{states}\ i}C_i\ e^{-\beta(\Delta_i-c_0)} = \sum_{\Delta=0}^{\infty}\overline{C_{\Delta}}\rho(\Delta) e^{-\beta(\Delta-c_0)}
\)
where $\overline{C_{\Delta}}$ is the average of the $C_i$ over all states with dimension $\Delta$ and $C_i>0$. We assume unitarity, a unique vacuum state with $C_\text{vac} =1$ and a gap. If $C=1$ and $c_0=c/12$ then $X=Z(\beta)$, the torus partition function. It will be useful to introduce a spectral representation for
$X$:

\(
X(\beta) = \int_0^{\infty} d\Delta\ Y(\Delta) e^{-\beta(\Delta-c_0)}.
\)
Then $Y(\Delta)$ is the inverse Laplace transform of $X$,
\(
\label{spectral}
Y(\Delta) = \int_{\gamma} \frac{d\beta}{2\pi i}\ X(\beta) e^{\beta(\Delta-c_0)}.
\)
When $C=1$, $Y$ is the spectral density $\sum_i \delta(\Delta-\Delta_i)$.

Let us further assume that $X$ transforms with weight
$w$ under the S-transformation $\beta\rightarrow 4\pi^2/\beta\equiv \beta'$,
\(
\label{modcov}
X(\beta) = \left(\frac{\beta}{2\pi}\right)^w X(4\pi^2/\beta)=\left(\frac{\beta}{\beta'}\right)^{w/2} X(\beta')\equiv
\alpha X'.
\)
Modular invariant quantities such as the partition function of course have $w=0$.

Modular covariance can be used to show that $X$ is approximated by its
contribution from appropriately defined light states. The argument
essentially follows \cite{Hartman:2014oaa}. First one splits $X$ up into contributions from light and heavy states
\(
X_L = \sum_{\Delta = 0}^{c_0+\epsilon}\overline{C_{\Delta}} e^{-\beta (\Delta-c_0)},\quad\quad
X_H = X-X_L
\)
for some $\epsilon>0$. Next, \eqref{modcov} implies

\(
\label{modcovsplit}
X_L + X_H = \alpha(X_L' + X_H').
\)
If $\beta>2\pi$ then $X_H$ is bounded by
$X_H'$:
\(
\label{hksineq}
X_H =\sum_{\Delta = c_0+\epsilon}^{\infty}C(\Delta) e^{-(\beta-\beta')
  (\Delta-c_0)}e^{-\beta' (\Delta-c_0)}\leq e^{(\beta'-\beta)\epsilon}
X_H' \equiv r X_H'.
\)
Following \cite{Hartman:2014oaa}, one can manipulate \eqref{modcovsplit} and \eqref{hksineq} into a bound on $X$ in terms of $X_L$:
\bea
\label{hks+}
 \log X_L &\leq& \log X \leq \log X_L -
\log\left(1-\frac{r}{\alpha}\right),
\eea
which is the result of \cite{Hartman:2014oaa} when $\alpha=1$. When $\beta<2\pi$, $X_L \rightarrow X'_L$ and $\alpha\rightarrow \alpha^{-1}$.

$X$ will be approximated by its contribution from light states when $\beta>2\pi$ if
$|\log\left(1-\frac{r}{\alpha}\right)|\ll \log X_L$, which is $O(c)$ for the quantities we study. The approximation
breaks down when
\(
\label{boundt}
1-r/\alpha = 1-e^{(\beta'-\beta)\epsilon -\frac{w}{2}\log\frac{\beta}{\beta'}}\sim O(e^{-c})
\)
i.e. when the exponent gets very close to zero. Whether or not it breaks down depends on the sign of $w$:
the first term in the exponent is always negative for $\beta>2\pi$,
but the sign of the second depends on $w$. If $w>0$ both terms are negative, so $\log X\approx \log X_L$ when $\beta>2\pi$. If $w<0$ the exponent crosses
zero as $\beta$ goes from $\infty$ to $2\pi$, at which point the upper bound 
becomes trivial. There are two cases: if $w$ is negative and $O(1)$ then $\epsilon$ must be adjusted to satisfy \eqref{boundt} for
all $\beta>2\pi$, but can remain $O(1)$. This value of $\epsilon$ is small compared to
$c_0\sim c$ when $c$ is large. However, if $w$ is negative and grows with $c$,  then we must either let $\epsilon$ grow with $c$ in
order to satisfy \eqref{boundt} (i.e. include ``heavy'' states in
$X_L$) or fix the definition of $X_L$, in which case $X\approx X_L$ will no longer hold
over a range of $\beta>2\pi$ that grows with $c$. Accordingly we limit ourselves to $X$ which have $w>M$ for some $O(1)$ $M<0$.

We will study the asymptotic behavior of the spectral
density $Y(\Delta)$ in \eqref{spectral} with $c_0\sim c$. There are multiple asymptotic
limits: $\Delta$ can be taken larger
than any other parameter in the problem, or taken to infinity with $\Delta/c$ fixed. Assuming the existence
of a thermodynamic description the integral
can be approximated via saddle point in either limit. When
$\Delta$ is taken larger than any other parameter the saddle is at $\beta\rightarrow
0$ since $X(\beta\rightarrow 0)\sim
e^{\beta'}$, while the saddle may be at
nonzero  $\beta$ if $\Delta/c$ is
held fixed.

First take
$\Delta\rightarrow \infty$ while keeping everything else
fixed:

\(
Y(\Delta\rightarrow \infty) \approx \int_{\gamma}\frac{d\beta}{2\pi i}\  X(\beta\rightarrow 0)
e^{\beta(\Delta-c_0)} = \int_{\gamma}\frac{d\beta}{2\pi i}\  \left(\frac{\beta}{2\pi}\right)^w
X(\beta'\rightarrow \infty)
e^{\beta(\Delta-c_0)}.
\)
We write $a \approx b$ to denote that the two quantities
have the same leading exponential behavior in the indicated limit,
i.e. $\frac{\log a}{\log b}\rightarrow 1$. If $X$ has an expansion of the form \eqref{boltz} then it is
dominated by the $\Delta=0$ term in the limit
$\beta\rightarrow \infty$:
\(
X(\beta\rightarrow \infty) \approx e^{\beta c_0}\equiv X_\text{vac}(\beta)
\)
and so
\(
Y(\Delta\rightarrow \infty) \approx \int_{\gamma}\frac{d\beta}{2\pi i}\  \left(\frac{\beta}{2\pi}\right)^w
X_\text{vac}(\beta')
e^{\beta(\Delta-c_0)}.
\)
Evaluating the integral then gives an approximate asymptotic formula
for $Y(\Delta)$. For example, if $X$ is the torus partition function
one obtains the Cardy formula\footnote{This is not precisely the
  correct expression, since the density of states is a sum of delta
  functions. The smooth expression~\eqref{cardya} arises from
  failure to account for the infinite nature of the sum
  in~\eqref{boltz}. However, ~\eqref{cardya} emerges after smearing over a small
  range of energies~\cite{Mukhametzhanov:2019pzy}.}

\(
\label{cardya}
\rho(\Delta\rightarrow \infty) \approx \int_{\gamma}\frac{d\beta}{2\pi i}\  e^{\frac{\pi^2 c}{3\beta}}
e^{\beta(\Delta-c/12)}\approx e^{2\pi\sqrt{\frac{c}{3}(\Delta-\frac{c}{12})}}.
\)
The latter expression can either be obtained via direct saddle analysis or by recognizing the integral as proportionate to the modified Bessel function $I_{\nu}(z)$, with argument $z\sim \sqrt{c\Delta}$, and expanding the Bessel function at large argument.

Now consider taking $c\rightarrow \infty$ with
$\Delta/c$ fixed. In this limit the inverse Laplace transform
\eqref{spectral} may be dominated by a saddle point at finite
$\beta$.  As long as the saddle is at $\beta>2\pi$ we can substitute $X\approx
X_L$ in \eqref{spectral}, but in general $X_L$ is not universal: it
could depend on all the light data of the theory. However, if $X_L \approx
X_\text{vac}$ in the limit, the asymptotic formula for
$Y(\Delta\rightarrow \infty)$ will have an extended regime of validity:

\(
\label{cardyb}
Y(c\rightarrow \infty,\Delta/c\text{ fixed}) \approx
\int_{\gamma}\frac{d\beta}{2\pi i}\  \left(\frac{\beta}{2\pi}\right)^w
X_\text{vac}(\beta')
e^{\beta(\Delta-c_0)}
\)
for all $\Delta$ such that the saddle is at $\beta>2\pi$.

Suppose we have a bulk argument that the
asymptotic formula for $Y$ remains valid in the extended
regime, for example the existence of thermodynamically stable black holes with $\Delta>c/6$ whose entropy is still given by \eqref{cardya}~\cite{Witten:1998zw}. Any CFT with a semiclassical bulk dual must then have
$X_L \approx X_\text{vac}$ at leading order. This in turn leads to a set of constraints on
the CFT data. For example, $Z(\beta>2\pi)\approx Z_\text{vac}$ if

\(
\sum_{\Delta=0}^{c/12+\epsilon} \rho(\Delta) e^{-\beta(\Delta-c/12)} \approx
e^{\beta c/12}.
\)
If the CFT
satisfies the sparseness constraint
\(
\rho(\Delta) \lesssim e^{2\pi \Delta}
\)
for all $\Delta \leq \frac{c}{12}+\epsilon$, the expression \eqref{cardya} for $\rho(\Delta\rightarrow\infty)$ remains valid for all $\Delta$ such that the saddle is at $\beta>2\pi$. Since the saddle is at
\(
\beta_\star = \sqrt{\frac{\pi^2 c}{3(\Delta-c/12)}}
\)
the asymptotic formula remains valid for all $\Delta >c/6$. While this example just recapitulates~\cite{Hartman:2014oaa} the
procedure leading to an extended regime of validity can be repeated
for any quantity $X$ with the properties above.

These arguments are readily generalized to independent left- and right-moving
temperatures, leading to asymptotic expressions $Y(\Delta, J)$
for averages over states with dimension $\Delta$ and spin $J$ which remain valid when
$\beta_\star,\bar{\beta}_\star>2\pi$. One simply
follows the argument in section 3.1 of \cite{Hartman:2014oaa}; introducing modular covariance just introduces factors of $\alpha$ as above.

In studying $\overline{C_{HHL}^2}$ and $\overline{C_{HHL}}$ we will encounter quantities for which $C$ has indefinite sign, so this approach will not work. In those cases we will use the results of~\cite{Kraus:2017kyl}, which takes a different approach to show that $X\approx X_L$ when $\beta>2\pi$ under certain additional assumptions.\footnote{\label{footnote}In addition to the HKS sparseness condition they assume factorization of what they call light correlators (between operators which have $\Delta<\Delta_c$, with $\Delta_c$ taken to infinity after performing the large $c$ expansion), subexponential
  growth of light correlators in medium states (which have $\Delta_c<\Delta<c/12+\epsilon$) and the existence of a
large $c$ expansion of the light contribution to the thermal correlator. We use the HKS definition of light ($\Delta<c/12+\epsilon$) and heavy except where noted otherwise.}

\section{Extended regimes}
\label{extended}
\subsection{$\overline{C_{HLL}^2}$: plane four-point function}
\label{sec:chll}

The plane four-point function encodes \cite{Das:2017cnv}  the
$\Delta_H\rightarrow\infty$ limit of $\overline{C_{HLL}^2}$ via the
``pillow'' representation of \cite{Maldacena:2015iua}, where the
four-point function is transformed to a new conformal frame in which
the operators are located at the corners of a pillow, $\mathcal{P} = T^2/\mathbb{Z}_2$. Taking all four operators to be identical, the relation between plane and pillow correlators is

\(
\langle O(0) O(z) O(1) O(\infty)\rangle_{\mathbb{C}}=\Lambda(q)\Lambda(\bar{q}) \langle O(0) O(\pi) O(\pi(\tau+1)) O(\pi\tau)\rangle_{\mathcal{P}} 
\)
where $q$ and $z$ are related via $q=e^{i\pi\tau}$, $\tau=i K(1-z)/K(z)$ with $K$ the elliptic integral of the first kind. The functions $\Lambda(q) = \theta_3(q)^{\frac{c}{2}-16h_O}[z(1-z)]^{\frac{c}{24}-2h_O}$ account for the conformal transformation of the operators, the Weyl
anomaly and the need to properly define the operators at
singular points of the transformation.

The pillow four-point function can be expressed in Boltzmann sum form by using the Hamiltonian to evolve the operators by $\pi\tau$:

\beq
\langle O(0) O(\pi) O(\pi(\tau+1)) O(\pi\tau)\rangle_{\mathcal{P}}&=&\bra{O(\pi)O(0)} q^{L_0-c/24} \bar{q}^{\bar{L}_0-c/24} \ket{O(\pi)O(0)}\cr
&\equiv&g(q,\bar q).
\eeq
We will take $\tau = \bar\tau^{\star} = \frac{i\beta}{2\pi}$, deferring 
unequal temperatures to the appendix. The transformation of $g$ under $\beta\rightarrow \beta'$, i.e. $\tau\rightarrow -1/\tau$, is determined by crossing symmetry: $z\rightarrow 1-z$ on the plane is modular symmetry $\tau\rightarrow -1/\tau$ on the pillow. The properties of $\Lambda$ imply
\(
g(\beta) = \left(\frac{\beta}{2\pi}\right)^{c/2-8\Delta_{O}} g(\beta')
\)
where $\beta'=\frac{4\pi^2}{\beta}$.

Inserting a complete set of states,\footnote{In \cite{Das:2017cnv} the sum over states is collected into a sum over
conformal families to obtain results for the OPE
coefficients averaged over heavy intermediate primaries. Here we will not do so and obtain results for the
OPE coefficients averaged
over all heavy intermediate states instead.}
\beq
g(\beta) &=&  \sum_{\text{states }i} C_{OOi}^2 16^{\Delta_i} e^{-\frac{\beta}{2}\left(\Delta_i-c/12\right)}\cr
&\equiv& \int_0^{\infty} d\Delta\ K(\Delta) 16^{\Delta}e^{-\frac{\beta}{2}\left(\Delta-c/12\right)}.
\eeq
The factor $16^{\Delta}$ arises from the difference between plane
and pillow OPE coefficients: the former multiply the terms in a $z$
expansion while the latter are defined by an expansion in
$q=\frac{z}{16}+O(z^2)$. The $C_{OOi}$ above are the OPE
coefficients on the plane.

$g$ takes the form of $X$ from section~\ref{generalities} with $w=c/2-8\Delta_{O}$. The corresponding spectral density is $K(\Delta)= \sum_i  C_{OOi}^2\delta(\Delta-\Delta_i)$, the OPE density. At infinite temperature,

\beq
g(\beta\rightarrow 0) &=& \left(\frac{\beta}{2\pi}\right)^{c/2-8\Delta_{O}} g(\beta'\rightarrow\infty)\cr
&\approx& \left(\frac{\beta}{2\pi}\right)^{c/2-8\Delta_{O}} e^{\frac{\pi^2 c}{6\beta}}.
\eeq
This determines the asymptotic OPE density:
\be
K(\Delta\rightarrow\infty) \approx 16^{-\Delta} \int_{\gamma} \frac{d\beta}{2\pi
  i}\  \left(\frac{\beta}{2\pi}\right)^{c/2-8\Delta_{O}} e^{\frac{\pi^2 c}{6\beta}} e^{\frac{\beta}{2}\left(\Delta-c/12\right)}
\ee
which has a saddle at $\beta_\star = \sqrt{\frac{\pi^2 c}{3(\Delta-c/12)}}$ as above, so
\be
K(\Delta\rightarrow\infty) \approx 16^{-\Delta}  e^{\pi\sqrt{\frac{c}{3}(\Delta-\frac{c}{12})}}
\ee
and
\(
\label{HLL}
\overline{C_{OO\Delta}^2}|_{\Delta\rightarrow\infty} = \frac{K(\Delta)}{\rho(\Delta)}\approx e^{-\pi\sqrt{\frac{c}{3}(\Delta-\frac{c}{12})}}\approx 16^{-\Delta}  e^{-S_\text{BH}(\Delta)/2}.
\)
This is the result of \cite{Das:2017cnv} for the squared OPE coefficient averaged over
primaries but with $c-1\rightarrow c$, as expected for a quantity averaged over
all states instead.

There is a bulk argument for this scaling~\cite{Das:2017cnv} but its status is
unclear. However, the fact that there are robust bulk arguments for both
$\overline{C_{HHL}^2}$ and $\rho(\Delta_H)$ is suggestive that one
should exist, perhaps a $2\rightarrow 2$ scattering process of light
particles in the bulk projected onto an intermediate black hole state,
whose amplitude is proportionate to $\overline{C_{HLL}^2}$. As the
mass of the intermediate black hole is taken to infinity the OPE
coefficient approaches its asymptotic regime of validity, and unless
the relevant physics changes significantly as the black hole goes from
$\Delta\sim c/6$ to $\Delta\gg c$ the asymptotic formula should extend. If this is correct then holographic field theories will be constrained to satisfy \eqref{HLL}, but in the absence of a robust bulk argument this is just a conjectural constraint.

Deriving the conditions under which \eqref{HLL} has such an extended regime of validity is straightforward. $g$ transforms with $w = c/2-8\Delta_{O}$, so $g\approx g_L$ at large $c$ when $\beta>2\pi$ provided $\Delta_{O} < c/16$, which we assume. In this case $g\approx g_\text{vac}$ for all $\beta>2\pi$ provided
\(
\overline{C_{OO\Delta}^2} \rho(\Delta)\lesssim e^{\pi\Delta}
\)
for all $\Delta \leq c/12+\epsilon$. Since the light OPE coefficients
are polynomial in $c$ in large $c$ CFTs, this is essentially just a
condition on the density of states, which is stronger than the HKS sparseness condition and excludes all permutation orbifolds \cite{Hartman:2014oaa,Haehl:2014yla,Belin:2014fna,Belin:2015hwa}. If this condition is obeyed, \eqref{HLL} remains valid for all such $O$ when $\Delta > c/6$. 

\subsection{$\overline{C_{HHL}^2}$: torus two-point function}
\label{sec:chhl}

The torus two-point function can be used to extract data on the asymptotics of heavy-heavy-light
OPE coefficients \cite{Brehm:2018ipf}. The starting point is the
thermal
autocorrelation function of a scalar on a spatial circle of length $L$:\footnote{Again we focus on
averages over all operators at dimensions $\Delta_{H_1},\Delta_{H_2}$ with any spin. In the appendix we obtain expressions for fixed spins $J_{H_1},J_{H_2}$ and spinning $O$.}

\be
 X(\beta,t)\equiv \tr\left[O(x=0,t)
                                                    O(0,0) e^{-\frac{2\pi\beta}{L}
                                                      \left(L_0+\bar{L}_0
                                                        - \frac{c}{12}\right)}
                                                    \right] = Z(\beta) \langle O(t) O(0)\rangle_{\beta}.  
\ee
In the $\beta\rightarrow \infty$ limit, the leading term is given by
\be
X(\beta\rightarrow\infty,t)\approx e^{\frac{\pi c\beta}{6L}}\frac{(-1)^{-\Delta_O}\left(\frac{\pi}{L}\right)^{2\Delta_O}}{\sin^{2\Delta_O}\left(\frac{\pi t}{L}\right)}
\ee
while the high temperature limit follows from a modular transformation,
\(
\label{hhl-hightemp}
X(\beta\rightarrow 0,t)
\approx e^{\frac{\pi c L}{6\beta}}
\frac{(-1)^{-\Delta_O}\left(\frac{\pi}{\beta}\right)^{2\Delta_O}}{\sinh^{2\Delta_O}\left(\frac{\pi
    t}{\beta}\right)}.
\)
We will take $L=2\pi$. Writing the torus two-point function as a sum over states,
\beq
\label{2pt}
X(\beta,t)&=& \sum_{i,j}\bra{i}O\ket{j}\bra{j}O\ket{i} e^{i(\Delta_i-\Delta_j)t} e^{-\beta\left(\Delta_i-\frac{c}{12}\right)}\cr%
&=& \int_0^{\infty} d\Delta \ \int_{-\infty}^{\infty}d\omega \ \ J(\Delta,\omega) e^{i\omega t} e^{-\beta\left(\Delta-\frac{c}{12}\right)}
\eeq
where the spectral density is
\(
J(\Delta,\omega) = \sum_{i,j} |\bra{i}O\ket{j}|^2
\delta(\Delta_i-\Delta)
\delta\left((\Delta_i-\Delta_j)-\omega\right).
\)
Eq. \eqref{2pt} can be inverted to solve for $J$:
\(
J(\Delta,\omega) = \int_{\gamma}
\frac{d\beta}{2\pi i}\int_{-\infty}^{\infty}
\frac{dt}{2\pi}e^{-i\omega t}
e^{\beta \left(\Delta-\frac{c}{12}\right)}X(\beta,t).
\)
As $\Delta\rightarrow \infty$ with $\omega$ fixed, the integral will
be dominated by its contribution from $\beta\rightarrow 0$. Using \eqref{hhl-hightemp}
and following the computation in
\cite{Brehm:2018ipf},
\be
  J(\Delta\rightarrow\infty,\omega\text{ fixed})\approx e^{2\pi\sqrt{\frac{c}{3}\left(\Delta_\text{avg} - \frac{c}{12}\right)}}\left|\Gamma\left(\Delta_O+\frac{i\omega}{\sqrt{12 \Delta_\text{avg}/c-1}}\right)\right|^2.
\ee
Here $\Delta_\text{avg} = \Delta+\omega/2$ and the exponential factor is the
Cardy density of states. Since
\(
J(\Delta,\omega) =\overline{C_{\Delta O (\Delta+\omega)}^2}
\rho(\Delta) \rho(\Delta+\omega)
\)
one obtains
\(
\label{asymp-hhl}
\overline{C_{\Delta O (\Delta+\omega)}^2}|_{\Delta \rightarrow \infty} \approx \ e^{-S_\text{BH}\left(\Delta_\text{avg}\right)}\left|\Gamma\left(\Delta_O+\frac{i\omega}{\sqrt{12\Delta_\text{avg}/c-1}}\right)\right|^2.
\)

This expression for the OPE density has a simple bulk interpretation
\cite{Brehm:2018ipf}. For $\Delta, \Delta' > \frac{c}{6}$, $C_{\Delta O
  \Delta '}^2$ is the probability for a transition between black hole
microstates while emitting a scalar. The probability of
transitioning between {\it any} two microstates and emitting a scalar is just
the emission probability, so the typical probability of transitioning into a
particular microstate and emitting a scalar is $e^{-S}$ times the
emission probability. Since the $\Gamma$ function factors in \eqref{asymp-hhl} give
precisely the probability of emission from a BTZ black hole as computed from the
quasinormal modes \cite{Maldacena:1997ih} the bulk calculation
matches the asymptotics \eqref{asymp-hhl} exactly.

The existence of a gravity argument for \eqref{asymp-hhl}
implies that it has an extended regime of validity in holographic CFTs, but the phases in \eqref{2pt} spoil the
positivity property necessary for
the argument of section~\ref{generalities}. However, one can derive the
extended regime using the results of~\cite{Kraus:2017kyl}. Consider the Witten diagram calculation of the CFT
two-point function at $\beta>2\pi$. The dominant geometry will be
thermal AdS so long as $O$ not too heavy (we can take it to be at most ``hefty'', i.e. with $\Delta_O\lesssim \varepsilon
c$ with $\varepsilon\ll 1$). At leading order in $1/c$ the Witten
diagrams that contribute correspond to free propagation between the
two boundary points, winding the thermal circle an arbitrary
number of times \cite{Hemming:2002kd}:
\beq
X(\beta>2\pi,t)|_{\text{bare}}&=&
e^{\frac{\pi^2 c}{3\beta}}\sum_{n=-\infty}^{\infty}\frac{(-1)^{-\Delta_O}\left(\frac{\pi}{\beta}\right)^{2\Delta_O}}{\sinh^{2\Delta_O}\left(\frac{\pi
      (t-2\pi n)}{\beta}\right)}
\cr
&\equiv&X_\text{vac}(\beta,t).
\eeq
At large but finite $c$ bulk
interactions dress the propagator and in principle we must sum over all intermediate
states -- including virtual black holes -- but when $\beta>2\pi$ in HKS-sparse large $c$
CFTs (with the mild additional assumptions stated above in footnote \ref{footnote}) heavy
intermediate states do not contribute~\cite{Kraus:2017kyl}. Since perturbative bulk interactions
are sub-exponential in $c$, \eqref{asymp-hhl} continues to give the leading
exponential behavior:
\beq
\label{xb2pi}
X(\beta>2\pi,t)&\approx&  X_\text{vac}(\beta,t).
\eeq
Eq. \eqref{cardyb} therefore implies that \eqref{asymp-hhl} remains valid for all $\Delta,\Delta' > c/6$: at finite $\beta>2\pi$, the spectral density is

\beq
J(c\rightarrow \infty,\Delta/c\text{ and }\omega\text{ fixed})
&\approx& \sum_{n=-\infty}^{\infty} J_n(\Delta,\omega)
  \eeq
where
 \beq
   J_n(\Delta,\omega) &\equiv& \int_{\gamma} \frac{d\beta}{2\pi
     i}\int_{-\infty}^{\infty} \frac{dt}{2\pi}\  X_\text{vac}(\beta,t-2\pi n)\cr
&=& e^{2\pi i n\omega} J(\Delta\rightarrow\infty,\omega\text{ fixed}),
\eeq
so
\be
 J(c\rightarrow \infty,\Delta/c\text{ and }\omega\text{ fixed})
\approx
    J(\Delta\rightarrow\infty,\omega\text{ fixed})\cdot \sum_{n=-\infty}^{\infty} \delta\left(\omega-n\right).
\ee
This implies that \eqref{asymp-hhl} continues to hold for $\Delta,
\Delta'>c/6$ whenever the CFT is HKS-sparse and obeys the mild
additional assumptions of \cite{Kraus:2017kyl}. The $\delta$ function reflects
the integer-spaced
spectrum of a free bulk field.

\subsection{$\overline{C_{HHL}}$: torus one-point function}
\label{sec:spots}

Next we consider $\overline{C_{HHL}}$ averaged over $H$, which is encoded by the torus
one-point function of a primary scalar
$O$~\cite{Kraus:2016nwo}. To recap, one writes this as a sum over states:

\(
\langle O\rangle_{\beta} = \sum_{\text{states }i}\bra{i}O\ket{i} e^{-\beta\left(\Delta_i-\frac{c}{12}\right)} =
\int_{0}^{\infty} d\Delta\ T(\Delta) e^{-\beta\left(\Delta-\frac{c}{12}\right)}
\)
where
\(
T(\Delta) = \sum_{\text{states }i} C_{iOi}
\delta(\Delta-\Delta_i)=\overline{C_{\Delta O\Delta}} \rho(\Delta).
\)
The S-transform of $\langle O\rangle_{\beta}$ is
\(
\langle O\rangle_{\beta} =  \left(\frac{\beta}{2\pi}\right)^{-\Delta_O}\langle
O\rangle_{\beta'}.
\)
As usual, the $\Delta_H\rightarrow
\infty$ asymptotics can be extracted (after a modular transformation) via inverse Laplace
transformation of the zero-temperature result
\(
\langle O\rangle_{\beta\rightarrow\infty} \approx \bra{\chi}O\ket{\chi}
e^{-\beta \left(\Delta_{\chi}-\frac{c}{12}\right)}
\)
where $\chi$ is the lightest operator with
$\bra{\chi}O\ket{\chi}\neq 0$. The modular property implies that
\(
\langle O\rangle_{\beta\rightarrow 0} \approx 
\bra{\chi}O\ket{\chi}\left(\frac{2\pi}{\beta}\right)^{\Delta_O}
e^{-\frac{4\pi^2}{\beta}\left(\Delta_{\chi}-\frac{c}{12}\right)}
\)
and so
\bea
T(\Delta\rightarrow \infty)&\approx&
\int_{\gamma}\frac{d\beta}{2\pi i}\ \langle O\rangle_{\beta\rightarrow 0} 
\cr
&\approx&
C_{\chi O\chi}e^{4\pi\sqrt{\left(\frac{c}{12}-\Delta_{\chi}\right)\left(\Delta-\frac{c}{12}\right)}}.
\eea
This leads to
an asymptotic expression for the average OPE coefficient \cite{Kraus:2016nwo},
\(
\overline{C_{\Delta O \Delta}}|_{\Delta\rightarrow\infty} \approx C_{\chi O \chi}\ e^{-\frac{\pi
    c}{3}\left(1-\sqrt{1-\frac{12
        \Delta_{\chi}}{c}}\right)\sqrt{\frac{12\Delta}{c}-1}}.
\)
If we take $c$ large with
$\Delta_{\chi}$ fixed
\(
\label{chlh}
\overline{C_{\Delta O \Delta}}|_{\Delta\rightarrow\infty} \approx C_{\chi O \chi}\  e^{-2\pi \Delta_{\chi}\sqrt{\frac{12\Delta}{c}-1}},
\)
which can be matched to a bulk calculation of the one-point function
via Witten diagrams on the BTZ background when
$\Delta_O\ll c$ \cite{Kraus:2016nwo}, so we again anticipate an extended
regime of validity in holographic CFTs.

As in section~\ref{sec:chhl}, one cannot simply run the modified HKS argument to
demonstrate extended validity since the $\overline{C_{\Delta O
    \Delta}}$ are not necessarily positive. However the arguments
of~\cite{Kraus:2017kyl} again allow us to proceed. Under the
same assumptions that lead to~\eqref{xb2pi}, the leading exponential
behavior of $\langle O\rangle_{\beta}$ is determined by a sum over light
states when $\beta>2\pi$, so~\eqref{chlh} will continue to hold for all
$\Delta>c/6$ provided\footnote{The definition of light states
  in~\cite{Kraus:2017kyl} is less inclusive than the one we have been
  using, which also includes what they call medium states. Since the
  contribution of medium states is subdominant we can extend their
  definition of light to $\Delta\leq c/12+\epsilon$ at no cost.}

\(
\langle O\rangle_{\beta,L} \equiv \sum_{\Delta=0}^{c/12+\epsilon}
\overline{C_{\Delta O \Delta}}\rho(\Delta)
  e^{-\beta\left(\Delta-\frac{c}{12}\right)}\approx C_{\chi O \chi}
  e^{-\beta \left(\Delta_{\chi}-\frac{c}{12}\right)}
  \)
  for all $\beta>2\pi$. This requires
  \(
  \rho(\Delta)\lesssim \frac{C_{\chi O \chi}}{\overline{C_{\Delta O
        \Delta}}} e^{2\pi(\Delta-\Delta_{\chi})}
  \)
  for all $\Delta\leq \frac{c}{12}+\epsilon$. This is the same condition that recently appeared
  in~\cite{Pal:2019yhz}. Since~\eqref{chlh} is demanded by the bulk this seems to
  be a constraint on holographic theories, but it is not appreciably
  stronger than the HKS condition: the ratio of OPE coefficients
  does not change the leading exponential behavior, and since the bulk
  argument assumes $\Delta_{\chi}\ll c$ the shift in the exponent is small.

The expression~\eqref{chlh} is not universal -- it depends on the
operator content and OPE coefficients -- but can be combined
with~\eqref{asymp-hhl} to obtain a bound on $C_{\chi O \chi}$ that depends only
on $\Delta_O$ and $\Delta_\chi$~\cite{Brehm:2018ipf}. However, if the
$\overline{C_{ijk}^2}$ do not vary wildly across heavy
states with fixed energy, their square root should approximate the
average of the unsquared coefficients, $\overline{C_{ijk}}$. This is
expected to be true in any chaotic theory. It is the average unsquared
coefficients that appear in the
discussion of vacuum block dominance in section~\ref{vbd}, but we will
assume that the average squared is approximated by the average of the squares and use the universal
expression~\eqref{asymp-hhl} in that discussion. Using \eqref{chlh} instead does not substantively change the
analysis.

\subsection{Density of primary states}

The asymptotic expression~\cite{Kraus:2016nwo} for the density of
primary states extends down to $\Delta>c/6$ under the ordinary HKS sparseness condition. One starts from the character
decomposition of the partition function:

\(
Z(\beta) = \sum_{h,\bar h} d_{h,\bar h} \chi_h(\beta) \chi_{\bar
  h}(\beta) \equiv \int d\Delta\
\rho_p(\Delta) e^{-\beta\left(\Delta-\frac{c-1}{12}\right)}|\eta\left(e^{-\beta}\right)|^{-2}
\)
where we used $\chi_{h\neq 0}(q) =
\frac{q^{h-(c-1)/24}}{\eta(q)}$ and took $\tau=\bar\tau^\star=\frac{i\beta}{2\pi}$. Inverting,
\(
\label{rhop}
\rho_p(\Delta) = \int_{\gamma} \frac{d\beta}{2\pi i}\ Z(\beta) |\eta(e^{-\beta})|^2
e^{\beta\left(\Delta-\frac{c-1}{12}\right)}.
\)
Under $\tau\rightarrow -1/\tau$,
\(
\eta(q) = (-i\tau)^{-1/2}
\eta(q') =\sqrt{\frac{2\pi}{\beta}}e^{-\frac{\pi^2}{6\beta}}\prod_{n=1}^{\infty}
(1-e^{-\frac{4\pi^2 n}{\beta}}) \approx \sqrt{\frac{2\pi}{\beta}}e^{-\frac{\pi^2}{6\beta}}
  \)
 as $\beta\rightarrow 0$. Therefore \cite{Kraus:2016nwo}
 \(
 \label{rhopasym}
\rho_p(\Delta\rightarrow \infty) \approx \int_{\gamma}
\frac{d\beta}{2\pi i}\
\left(\frac{2\pi}{\beta}\right) e^{\frac{\pi^2(c-1)}{3\beta}}
e^{\beta\left(\Delta-\frac{c-1}{12}\right)}\approx e^{2\pi\sqrt{\frac{c-1}{3}\left(\Delta-\frac{c-1}{12}\right)}},
\)
which is the Cardy formula with $c\rightarrow c-1$.

The shift $c\rightarrow c-1$ can be understood in the bulk as arising
from the Casimir energy of boundary
gravitons~\cite{Maxfield:2019hdt}. Since we are resumming descendants
in the bulk the resulting black hole entropy counts
the number of primaries on the CFT side, so we expect that~\eqref{rhopasym} remains valid in the extended regime. Assuming
HKS sparseness at large $c$ we still have $Z(\beta>2\pi)\approx
Z_\text{vac}(\beta)$, but the $\eta$ functions cannot be
approximated by their asymptotic form and a bit more work is
needed. The key step is the pentagonal number theorem,

\(
\prod_{n=1}^{\infty}(1-x^n) = \sum_{k=-\infty}^{\infty}(-1)^{k}
x^{\frac{3k^2-k}{2}}.
\)
Eq. \eqref{rhop} then reads

\beq
\rho_p(\Delta\rightarrow\infty,\Delta/c\text{ fixed}) &=&
\sum_{k=-\infty}^{\infty} (-1)^{k+k'} \int_{\gamma} \frac{d\beta}{2\pi
  i}\ \left(\frac{2\pi}{\beta}\right) e^{\frac{\pi^2(c-1-6f(k,k'))}{3\beta}}
e^{\beta(\Delta-(c-1)/12)}\cr
&=& 2\pi \sum_{k=-\infty}^{\infty} (-1)^{k+k'} I_0\left(2\pi\sqrt{\frac{c-1-6f(k,k')}{3}\left(\Delta-\frac{c-1}{12}\right)}\right)\cr
&\equiv& \sum_{k=-\infty}^{\infty} Q(k,k')
\eeq
where
\(
f(k,k') = 3(k^2+{k'}^2)-k-k'
\)
and $I_0$ is a modified Bessel function.

The behavior of $Q(k,k')$ is
approximately constant over three regions of the $k,k'$ plane:
$f(k,k')\ll c$, $f(k,k')\sim c$ and $f(k,k')\gg c$. When $f(k,k')\ll c$ it is obvious that $Q(k,k')\approx Q(0,0)$,
and since the number of $k,k'$ in this region grows sub-exponentially
with~$c$ the contribution of this entire region is $\approx
Q(0,0)$. When $f(k,k')\sim c$,
\(
Q(k,k') = 2\pi (-1)^{k+k'}
I_0\left(2\pi\sqrt{N\left(\Delta-\frac{c-1}{12}\right)}\right)
\)
where $N$ is an $O(1)$ number. Since $I_0(z)\sim z^{-1/2} e^{z}$ at large $z$,
$Q(k,k')\ll Q(0,0)$ in this region. Finally,
when $f(k,k')\gg c$, $Q(k,k')$ becomes an ordinary Bessel
function since $I_{0}(iz) =
J_{0}(-z)$. Since $J_0(z\rightarrow -\infty)\sim
(-z)^{-1/2}\cos(z+\pi/4)$ the contribution from large $k,k'$ is also subleading:
\(
\sum_{k=-\infty}^{\infty} Q(k,k')\approx Q(0,0)=I_0\left(2\pi\sqrt{\frac{c-1}{3}\left(\Delta-\frac{c-1}{12}\right)}\right).
\)
Using the asymptotic behavior of $I_0$ we then have
\(
\rho_p(\Delta\rightarrow \infty,\Delta/c\text{ fixed}) \approx
e^{2\pi\sqrt{\frac{c-1}{3}\left(\Delta-\frac{c-1}{12}\right)}}.
\)
In HKS-sparse theories the asymptotic form of $\rho_p$ thus extends to all $\Delta>c/6$.


\section{Further remarks}
\label{further}
This section is more technical and less conclusive than the rest
of the paper, but will be useful for the discussion of vacuum block
dominance in section~\ref{vbd}.

\subsection{$\overline{C_{HHH}^2}$: genus two partition function}
\label{sec:chhh}

The asymptotic form of $\overline{C_{HHH}^2}$ was obtained in~\cite{Cardy:2017qhl} by studying appropriate twist correlation
functions on the plane:

\(
\label{zmn}
Z_{m,n}\equiv \langle \prod_{k=1}^m \sigma_n(u_k) \bar{\sigma}_n(v_k)\rangle.
\)
Focusing for simplicity on the case of
equal dimensions $\Delta_H$, the asymptotics can be extracted from the
OPE singularities of \eqref{zmn} with 
$m=2,n=3$, or with $m=3,n=2$ and a $\Z_3$ symmetry relating the
$(u_k,v_k)$. Each of these correlation functions corresponds to the
partition function on a $\Z_3$-symmetric genus two Riemann surface in a
particular conformal frame, with an explicit formula relating $Z_{2,3}$ to
$Z_{3,2}$ given in~\cite{Cardy:2017qhl}. We will focus on
$\Z_3$-symmetric twist-2 six point functions, taking
\(
u_k = e^{i(2\pi k+\theta)/3},\quad v_k =e^{i(2\pi k-\theta)/3}.
\)
The parameter $\theta$ is related to the cross-ratio of the twist-2
four point function via $z=\cos^2\frac{\theta}{2}$.

The
$\Z_3$ symmetry reduces the three moduli of the $g=2$ surface to a
single modulus $\tau$, in terms of which the period matrix is~\cite{Calabrese:2009ez}

\(
\Omega = \frac{1}{\sqrt{3}} 
  \left( \begin{array}{cc}
   2 & -1 \\
   -1 & 2 \\
  \end{array}  \right)\tau,
\)
where $\tau$ can be expressed in terms of $z$ as a ratio of
hypergeometrics, just as in the genus one case from section~\ref{sec:chll}. The period matrix
transforms under genus 2 modular transformations, which are elements
of $Sp(4,\Z)$. Crossing symmetry of the
four-point function sends $\tau\rightarrow -1/\tau$: the twist correlators are invariant under $z\rightarrow 1-z$ or $\theta\rightarrow \pi-\theta$. The strategy for obtaining the asymptotics of $\overline{C_{HHH}^2}$ is to use the twist OPE:

\(
\sigma_2(x)\sigma_2(0) = \sum_{m} x^{h_m-c/8} \bar{x}^{\bar{h}_m-c/8}
C_{\sigma_2\sigma_2 m_{(ij)}} O_{m_{(ij)}}(0)
\)
where the $O_m$ are untwisted operators in the orbifold CFT,
\(
O_{m_{(ij)}} =O_i^{(1)}\otimes_\text{sym} O_j^{(2)}.
\)
Twist OPE coefficients are related to plane correlation functions by
the appropriate conformal transformation~\cite{Calabrese:2010he}. For primary $m$,
$C_{\sigma\sigma m_{(ij)}} \propto \delta_{ij}$ and so their
contribution to \eqref{zmn} is proportionate to the square of their OPE
coefficients with the operators coming from the other two twist OPEs. For descendants the OPE
coefficients will mix states within each conformal family, but since their
contributions are determined kinematically from the primary ones they can be collected into blocks:
\(
\label{hhhspec}
\langle\prod_{k=1}^m \sigma_n(u_k) \bar{\sigma}_n(v_k)\rangle =
\sum_{\text{primaries }ijk} C_{ijk}^2
|\mathcal{F}_{g=2}(h_{ijk},\theta)|^2 = \int d^3h\ d^3\bar{h}\ P(h_{123},\bar{h}_{123})
\)
where as usual a spectral representation
\(
P(h_{123},\bar{h}_{123}) = \overline{C_{(h_1,\bar{h}_1),
    (h_2,\bar{h}_2), (h_3,\bar{h}_3)}^2}
\rho_p(h_1,\bar{h}_1)\rho_p(h_2,\bar{h}_2)\rho_p(h_3,\bar{h}_3) |\mathcal{F}_{g=2}(h_{123},\theta)|^2
\)
was introduced.

In~\cite{Cardy:2017qhl} the blocks $\mathcal{F}_{g=2}$ are computed at
large $c$ (so that the monodromy method can be used) with
$h\gg c$ (so that the monodromy equation can be
solved, via WKB approximation). Referring to~\cite{Cardy:2017qhl} for the details, one can
take $\theta\rightarrow 0$ and invert \eqref{hhhspec} to obtain
\(
\label{hhhasymp}
\overline{C_{(h_1,\bar{h}_1),
    (h_2,\bar{h}_2), (h_3,\bar{h}_3)}^2}|_{\text{all }h,\bar
  h\rightarrow \infty}\approx
e^{-3\pi\left(\sqrt{\frac{c}{6}h}+\sqrt{\frac{c}{6}\bar
      h}\right)}=\rho_\text{BH}(h,\bar h)^{-3/2}.
\)
No bulk argument for~\eqref{hhhasymp} has been proposed, nor have we come up with one. We have no reason to expect it to hold as $c\rightarrow\infty$
with $h/c$ fixed. However the computation has the same flavor as those
in section~\ref{extended} and it is natural to ask if the same methods
can be applied here. Unfortunately, one quickly runs into a series
of obstacles, each interesting but unsurmounted, and ultimately we
will only be able to give a rough argument.

Extending the computation of~\cite{Cardy:2017qhl} to $h\sim
O(c)$ requires knowledge of the blocks in that limit, where the
WKB approximation breaks down. One obtains a differential equation of Heun type for the
accessory function. This is analogous to (but harder than) the
unsolved problem of obtaining a closed-form expression for the blocks on the
plane when $h\sim O(c)$.

The blocks make no appearance if we choose not to resum the
contributions to $Z_{3,2}$ from each conformal family, but then the correlator~\eqref{zmn} does not take the form of a sum
over squared OPE coefficients since the descendant OPE coefficients
mix states within conformal families. As a last resort we can consider the genus two partition function in the
plumbing frame (reviewed in~\cite{Cardy:2017qhl,Cho:2017oxl}) in which the surface is represented
by two spheres connected by three cylinders of equal height $\ell$ and unit radius, which are
glued to the spheres at $0,1,\infty$. Inserting
complete sets of states at $0,1,\infty$ on one of the spheres,
\(
Z_{g=2,\text{plumb}} = \sum_{\text{states }i,j,k} C_{ijk}^2
e^{-\ell(\Delta_i+\Delta_j+\Delta_k-c/4)}.
\)
Unfortunately, the relationship between $\ell$ and $\tau$ is not
clear~\cite{Calabrese:2009ez}. Furthermore, the action of $\tau\rightarrow -1/\tau$ on
this partition function is not known: no closed form expression is
available for the conformal anomaly relating the partition
functions in the plumbing and twist operator frames (though it has
been obtained in a cross-ratio expansion \cite{Cho:2017fzo}). However,
we can make some educated guesses and see how far they take us.

First we conjecture that $\ell=\beta/2$, where $\tau=\frac{i\beta}{2\pi}$. This can be motivated by
thinking about the plumbing construction of the torus: traversing one of the bridges only takes you halfway around
the loop. Second we conjecture that the partition function transforms as
a modular form with non-negative weight. This is true of the genus 2 partition
function in other conformal
frames~\cite{Mathur:1988xc,Witten:2007kt,Gaiotto:2007xh}, including
the twist frame. Under these assumptions one
can rederive~\eqref{hhhasymp} using the plumbing frame:
\(
Z_{g=2,\text{plumb}}(\beta)  = \int d^3\Delta\ 
g(\Delta_{123}) e^{-\frac{\beta}{2}\left(\Delta_1+\Delta_2+\Delta_3-c/4\right)}
\)
where
\(
g(\Delta_{123}) = \overline{C^2_{\Delta_1\Delta_2\Delta_3}}\rho(\Delta_1)\rho(\Delta_2)\rho(\Delta_3)
\)
and the OPE coefficient is averaged over all states with
dimensions $\Delta_1$, $\Delta_2$ and $\Delta_3$. We also have
\(
Z_{g=2,\text{plumb}}(\beta\rightarrow \infty)\approx e^{\frac{\beta
    c}{8}}.
\)
Changing variables
to $E=\Delta_1+\Delta_2+\Delta_3$, $\omega_1 =
\Delta_1+\Delta_2-\Delta_3$ and $\omega_2 =
\Delta_1-\Delta_2+\Delta_3$,
\(
Z_{g=2,\text{plumb}}(\beta)  = \int_{0}^{\infty} dE\ 
K(E) e^{-\frac{\beta}{2}\left(E-c/4\right)}
\)
where $K(E) = \sum_{ijk} C_{ijk}^2
\delta(E-\Delta_i-\Delta_j-\Delta_k)$ up to an $O(1)$ multiplicative
factor. This leads to an asymptotic expression for the OPE density:
\(
K(E\rightarrow\infty) \approx \int_{\gamma} \frac{d\beta}{4\pi i}
e^{\frac{\beta}{2}\left(E-c/4\right)}
Z_{g=2,\text{plumb}}(\beta\rightarrow 0) = \int_{\gamma}\frac{d\beta}{4\pi i}\left(\frac{\beta}{2\pi}\right)^w
e^{\frac{\beta}{2}\left(E-c/4\right)}
Z_{g=2,\text{plumb}}(\beta'\rightarrow \infty).
\)
Evaluating on the saddle, one finds
\(
\label{asymphhh}
\overline{C_{\Delta\Delta\Delta}^2}|_{\Delta\rightarrow\infty} \approx
\frac{K(E\rightarrow\infty)}{\rho(\Delta\rightarrow\infty)^3}\approx
e^{-3\pi\sqrt{\frac{c}{3}\left(\Delta-\frac{c}{12}\right)}} =
\rho_\text{BH}(\Delta)^{-3/2}.
\)

When $\Delta\sim O(c)$ the saddle point is at finite $\beta$ and we
must proceed as in section~\ref{generalities}. The argument of that section is readily
extended to triple sums to show that
\(
Z_{g=2,\text{plumb}}\approx Z_{g=2,\text{plumb},L} 
\)
when $\beta>2\pi$, where the quantity on the RHS contains the
contributions of states $i,j,k$ with at least one of $\Delta_i,\Delta_j,\Delta_k \leq
c/12+\epsilon$. Next we ask when $Z_{g=2,\text{plumb},L}(\beta>2\pi)\approx
  Z_{g=2,\text{plumb,vac}}$. There are three
  types of contributions to $Z_{g=2,\text{plumb},L}$: from three
  light operators, from two light
  operators and one heavy, and from one light and two heavy. If we
  assume that $\rho(\Delta)~\lesssim~e^{\pi\Delta}$ then we can use
  \eqref{hlleq} and \eqref{hhleq} to conclude that the $HLL$ and
  $HHL$ contributions do not affect the leading exponential
  behavior. However, the $LLL$ OPE coefficients grow
  exponentially when the light operators are multitraces. Counting
  contractions and using Stirling's formula, the
  most dangerous kind take the
  form~\cite{Belin:2017nze}\footnote{There are also multitrace contributions of
    the form $C_{:\tr \phi_1^{K} :\ :\tr \phi_2^{K}:\ :\tr
 \phi_3^{K}:}$ but the same counting argument shows that these are $O(1)$.}

 \(
 C_{:\tr \phi^{K_1} :\ :\tr \phi^{K_2}:\ :\tr
 \phi^{K_3}:} \approx 2^{\Delta/\Delta_{\phi}}
 \)
 where $\Delta = (K_1+K_2+K_3)\Delta_{\phi}$. The largest contribution
 comes from multitraces composed of the lightest non-identity operator in the
 theory, with dimension $\Delta_\text{min}$, and so $Z_{g=2,\text{plumb},L}(\beta>~2\pi)\approx
  Z_{g=2,\text{plumb,vac}}$ when
  \(
 \rho(\Delta) \lesssim e^{(2\pi - \frac{\log 2}{\Delta_\text{min}})\Delta}.
 \)
This becomes stronger than $\rho(\Delta)\lesssim e^{\pi\Delta}$ when
$\Delta_\text{min}\leq .2206\dots$ and requires essentially no light
states when $\Delta_\text{min}\leq .1103\dots$. Under this condition
\eqref{asymphhh} will continue to hold for all $\Delta>c/6$.
  
  It is intriguing that one can recover~\eqref{hhhasymp} and extend its
  regime of validity by making two plausible conjectures, but
  the argument is neither rigorous nor motivated by the bulk.

\subsection{Averages over primary states}
\label{primav}

We showed in section~\ref{generalities} that the expression \eqref{rhopasym} for the density of primary states has an
extended regime of validity in sparse theories at large $c$,
as implied by the bulk. One might expect a
similar story for the averaged primary OPE coefficients, but the unknown
structure of the conformal blocks when $\Delta\sim c$ prevents an
analysis along the lines of section~\ref{generalities}. As an example
consider the pillow four-point function, with the contributions from each
conformal family collected into blocks~\cite{Maldacena:2015iua}:

\beq
\label{chllp}
g(\beta)&=&\sum_{\text{primaries }\alpha} C_{OO\alpha}^2
V(h_{\alpha},\beta) \bar{V}(\bar{h}_{\alpha},\beta)\cr
&\equiv&\int_0^{\infty} dh d\bar{h}\ K_p(h,\bar{h}) V(h,\beta) \bar{V}(\bar h,\beta),
\eeq
where the conformal blocks are
\(
V(h,q) = 16^{h-c/24}q^{h-(c-1)/24}\eta(q^2)^{-1/2}H(h,q)
\)
and the $H(h,q)$ are the standard Zamolodchikov $H$-functions, which can be
computed recursively \cite{Zamolodchikov:1985ie,Zamolodchikov1987}. Since
$H(h\rightarrow\infty,q)\approx 1$, eq.~\eqref{chllp} expresses
$g(\beta\rightarrow 0)$ as a Laplace transform. This can be
inverted to obtain $K_p(\Delta\rightarrow\infty)$ and thus an
expression for the averaged primary OPE coefficients \cite{Das:2017cnv}

\(
\label{chllpres}
\overline{C^2_{OO\Delta}}|_{\Delta\rightarrow \infty} \approx
16^{-\Delta} e^{-\pi\sqrt{\frac{c-1}{3}(\Delta-\frac{c-1}{12})}}
\)
which is the result \eqref{HLL} for the average over all states with
$c\rightarrow c-1$. Since
\(
V(h,q)\bar{V}(\bar{h},q) \rightarrow (16q)^{\Delta}
\)
as $h,\bar h\rightarrow \infty$, the $16^{-\Delta}$ just reflects the
convergence of the OPE.

When $h\sim c$ it is no longer true that $H(h,q)\approx 1$, so 
\eqref{chllp} is not a Laplace transform unless the $H$ function exponentiates, i.e.
$H(h,q)\approx q^{ah}$ for some $a$. If this is the case then one can
again invert to obtain~\eqref{chllpres} for all $\Delta> c/6$ under
the conditions of section~\ref{sec:chll}. We expect this to be true, and preliminary
numerics~\cite{shouvik_note}
suggest that this is indeed the case, but the lack of a closed form
expression prevents us from saying more.

The situation is similar for $\overline{C_{HHL}^2}$ and
$\overline{C_{HHL}}$: when the sums over states are collected into sums over primaries, the integrals over the corresponding spectral densities only take the form of a
Laplace transform if the torus blocks
exponentiate when $h\sim c$. The necessary analysis is left for future work.

\section{Vacuum block dominance}
\label{vbd}

Gauge-gravity duality suggests that heavy-light correlation
functions are well-approximated by the contribution from the vacuum
Virasoro block in holographic CFTs:\footnote{If there are any conserved currents their
  Virasoro blocks may
  also contribute at leading order, corresponding to the effects of
  gauge fields in the bulk. In this case
  the appropriate notion is dominance of the
  vacuum block of the extended chiral algebra.} 
the entanglement entropy thus computed agrees with the RT formula
\cite{Asplund:2014coa}, while in the chaos regime the vacuum block exhibits the
Lyapunov behavior of particles near the horizon
\cite{Shenker:2013pqa,Anous:2019yku}. This is closely connected to the eigenstate
thermalization hypothesis, since the vacuum block contribution matches
the thermal expectation value at the appropriate temperature~\cite{Fitzpatrick:2014vua}.

The original motivation for this work was to find a set of conditions
on large $c$ CFTs under which the vacuum block
dominates these correlation functions. While we cannot fully answer this question, in
this section we outline progress that can be made using the results
above and highlight the missing ingredients. Restricting to Euclidean four-point functions for simplicity, the correlator can be expanded as

\beq
\label{4pt}
\langle O_H(0) O_L(1) O_L(z) O_H(\infty)\rangle &=& \sum_{\text{primaries }p} C_{LLp} C_{HHp} |\mathcal{F}(h_L,h_H,h_p,c,z')|^2\cr
&=& \sum_{h_p,\bar{h}_p} \overline{C_{LL(h_p,\bar{h}_p)} C_{HH(h_p,\bar{h}_p)}} \rho_p(h_p,\bar{h}_p) |\mathcal{F}(h_L,h_H,h_p,c,z')|^2\cr
&&
\eeq
where $z'\equiv 1-z$. The average is over all primaries with weights $(h_p,
\bar{h}_p)$ and we take $(h_H,\bar{h}_H)\gtrsim c/12$, $(h_L,\bar{h}_L)\sim O(1)$. The sum can be split into three regions:
\beq
\text{Region I:}&\quad\quad&h_p,\bar{h}_p\ll c\cr
\text{Region II:}&\quad\quad&h_p\text{ and/or }\bar{h}_p \sim c\text{ but neither}\gg c\cr
\text{Region III:}&\quad\quad&h_p\text{ and/or }\bar{h}_p\gg c.
\eeq
The basic ingredients are the OPE density and conformal blocks in each region.

The blocks are known in regions I~\cite{Fitzpatrick:2015zha} and
III~\cite{Harlow:2011ny} but must be computed recursively in most of region II~\cite{Zamolodchikov:1985ie,Zamolodchikov1987,Fitzpatrick:2014vua}. Meanwhile, the OPE coefficients are $\overline{C_{LL(h_p,\bar{h}_p)} C_{HH(h_p,\bar{h}_p)}}$, which we have not found tools to study directly. We will have to make some assumptions. Under the assumption that the OPE coefficients are not wildly varying, and that the light and heavy factors are statistically independent, we can approximate

\(
\overline{C_{LL(h_p,\bar{h}_p)} C_{HH(h_p,\bar{h}_p)}}\approx \sqrt{\overline{C_{LL(h_p,\bar{h}_p)}^2}} \sqrt{\overline{C_{HH(h_p,\bar{h}_p)}^2}}\equiv \overline{C_{h_p,\bar{h}_p}}
\)
so that 
\beq
\langle O_H(0) O_L(1) O_L(z) O_H(\infty)\rangle &\approx& \sum_{h_p,\bar{h}_p} \overline{C_{h_p,\bar{h}_p}}\rho_p(h_p,\bar{h}_p) |\mathcal{F}(h_L,h_H,h_p,c,z')|^2.
\eeq
Next we assume that the $\Delta> c/6$ OPE coefficients
averaged over all primaries take the form of the corresponding OPE
coefficients averaged over all states with the replacement
$c\rightarrow c-1$, as we showed for the $\Delta> c/6$
density of primaries. We also assume that $O_H$ is typical in the
sense that $\overline{C_{HH(h_p,\bar{h}_p)}^2}$ averaged over $O_p$
is well-approximated by $\overline{C_{HH(h_p,\bar{h}_p)}^2}$ averaged over $O_H$
as well. Finally we assume all sparseness conditions necessary for the
asymptotic expressions \eqref{hlleq}, \eqref{hhleq}, \eqref{hhheq} to remain valid in the extended regime. Under these assumptions we know the leading form of $C(h_p,\bar{h}_p)$ across most of the spectrum.

The contribution
 of region I is
 \beq
 \label{regioni}
\langle O_H(0) O_L(1) O_L(z) O_H(\infty)\rangle_\text{I} &=& 
|\mathcal{F}_\text{vac}|^2\left[1+\sum_{h_p,\bar{h}_p \ll c}
  \overline{C_{LL(h_p,\bar{h}_p)} C_{HH(h_p,\bar{h}_p)}}
  \rho_p(h_p,\bar{h}_p) \frac{\mathcal{F}_{h_p}\bar{\mathcal{F}}_{\bar{h}_p}}{|\mathcal{F}_\text{vac}|^2}\right]. \cr
&&
\eeq
This
 will be approximated by the vacuum block if the sum
 inside the brackets is much smaller than 1. In this region the blocks
 globalize, eq. (3.29) of \cite{Fitzpatrick:2015zha}. From that
 expression it follows that
$\frac{\mathcal{F}_{h_p}}{\mathcal{F}_\text{vac}}\approx
1$ at fixed $z$. As for the OPE density, the light-light-light coefficients are at most polynomial in~$c$,
while under our assumptions

\(
\overline{C_{HH(h_p,\bar{h}_p)}} \rho_p(h_p,\bar{h}_p) \approx
\rho_{BH}(\Delta_H)^{-1/2} \rho_p(\Delta_p) \lesssim
\rho_{BH}(\Delta_H)^{-1/2} \times e^{\pi\Delta_p}\sim e^{-c}.
\)
Thus we conclude that the vacuum block gives the leading exponential behavior in region I.

The contribution from very heavy intermediate states (region III) can be estimated without our results. Under the assumptions above, when $h_p$ or
$\bar{h}_p\gg c$ the OPE density is
\(
\overline{C_{LL(h_p,\bar{h}_p)} C_{HH(h_p,\bar{h}_p)}}
\rho_p(h_p,\bar{h}_p)\approx 16^{-\Delta_p}\rho_{BH}(\Delta_p)^{-1/2}
\rho_{BH}(\Delta_p)\sim 16^{-\Delta_p} e^{\sqrt{c\Delta_p}}
\)
while the block scales as
\(
\label{qdelta}
\lim_{h_p,\bar{h}_p\rightarrow \infty}
|\mathcal{F}(h_L,h_H,h_p,c,z')|^2\approx (16q')^{\Delta_p}.
\)
Their product must be summed over all $\Delta_p$ in region III:
\beq
 \label{regioniii}
\langle O_H(0) O_L(1) O_L(z) O_H(\infty)\rangle_\text{III} &\approx& 
\int_{\Delta^\text{III}_\text{min}}^{\infty} d\Delta\ {q'}^{\Delta} =
-\frac{{q'}^{\Delta^\text{III}_\text{min}}}{\log q'}.
\eeq
Here $\Delta^\text{III}_\text{min}\gg c$ is the smallest operator dimension in region III. This contribution is therefore exponentially smaller than the vacuum block contribution unless $q'\rightarrow 1$, i.e. $z\rightarrow 0$ or $\infty$.

It is much more difficult to bound the contribution of region II, where the
blocks must be computed recursively. Under our assumptions, when
$\Delta_p>c/6$ we have
$\overline{C_{LL(h_p,\bar{h}_p)}}\approx e^{-S_\text{BH}(\Delta_p)/4}$, but our expression for
$\overline{C_{HH(h_p,\bar{h}_p)}}$
is only conjectural. If the conjectured
form is correct the OPE density in this region is roughly $O(1)$, so the
contribution from states in region II with $\Delta_p>c/6$ will be subleading when the blocks provide exponential
suppression. When $\Delta_p<c/6$
we have no estimate for the OPE density.

We must also consider the crossover between the regions. Between
regions I and II there are ``hefty'' operators, which have $\Delta_p=\varepsilon c$ with $\varepsilon\ll 1$. The conformal block for hefty exchange is \cite{Fitzpatrick:2014vua}
\(
 \mathcal{F}(h_L,h_H,h_p=\varepsilon c,c,z') \approx(4\rho')^{h_p}
\)
to leading order in $\varepsilon$, where
\(
\rho(z) = \frac{z}{(\sqrt{1-z}+1)^2} = \left(\frac{\theta_2(q^2)}{\theta_3(q^2)}\right)^2
\)
is the $\rho$ variable of the bootstrap literature
\cite{Pappadopulo:2012jk}. The contribution from hefty
intermediate states scales like 
\(
\overline{C_{LL(\text{hefty})}
  C_{HH(\text{hefty})}} \rho_\text{hefty}
|\mathcal{F}_\text{hefty}|^2\lesssim  e^{-S_\text{BH}(\Delta_H)/2} e^{\pi \varepsilon c}  (4\rho')^{\varepsilon
  c} \sim e^{-c}
\)
and so is subleading to the vacuum block. However the crossover between regions II and III is murky
since a closed-form expression for the blocks is not known.

A more complete study of vacuum block dominance would require careful examination of
the validity of the assumptions above, \eqref{hhheq}
in the extended regime, the OPE density when $\Delta \lesssim c/6$ and the blocks when $\Delta\sim c$. All this is left for future work.

\section*{Acknowledgements}
I am grateful for extensive discussions with Shouvik Datta, Allic Sivaramakrishnan and especially Per Kraus, who also provided comments on the draft.

\appendix


\section{Results at fixed spin}

\subsection*{$\overline{C_{HLL}^2}$: plane four-point function}

In this section we generalize the analysis of \cite{Das:2017cnv}
to obtain asymptotic results for $\overline{C_{H LL}^2}$ averaged over
all operators with dimension $\Delta_{H}$ and spin $J_{H}$. The logic
is the same as in section~\ref{sec:chll}, but we take independent left and right inverse
temperatures $\beta,\bar\beta$, with $\tau=\frac{i\beta}{2\pi}$ and $\bar\tau = \frac{i\bar\beta}{2\pi}$:

\beq
g(q,\bar{q}) &=&\sum_{\text{states i}} C_{OOi}^2 e^{-\frac{\beta}{2}\left(h_i-\frac{c}{24}\right)-\frac{\bar\beta}{2}\left(\bar{h}_i-\frac{c}{24}\right)}\cr
&\equiv& \int_0^{\infty} dh d\bar{h}\ K(h,\bar h)e^{-\frac{\beta}{2}\left(h-\frac{c}{24}\right)-\frac{\bar\beta}{2}\left(\bar h-\frac{c}{24}\right)}.
\eeq
Under $\tau\rightarrow -1/\tau$ this transforms as
\(
g(q,\bar{q}) = (-i\tau) ^{c/4-8h_{O}}(-i\bar{\tau})^{c/4-8\bar{h}_{O}} g(q',\bar{q}')
\)
where $q'=e^{-i\pi/\tau}$.

At high temperatures,

\beq
g(\beta\rightarrow 0,\bar\beta\rightarrow 0) &=& \left(\frac{\beta}{2\pi}\right)^{c/4-8h_{O}}\left(\frac{\bar\beta}{2\pi}\right)^{c/4-8\bar{h}_{O}} g(4\pi^2/\bar\beta,4\pi^2/\beta\rightarrow\infty)\cr
&\approx& \left(\frac{\beta}{2\pi}\right)^{c/4-8h_{O}}\left(\frac{\bar\beta}{2\pi}\right)^{c/4-8\bar{h}_{O}}  e^{\frac{\pi^2 c}{12\beta}+\frac{\pi^2 c}{12\bar\beta}}.
\eeq
This determines the asymptotic OPE density:
\beq
K(h,\bar h\rightarrow\infty) \approx \int_{\gamma\times\bar\gamma} \frac{d\beta}{4\pi i}\frac{d\bar\beta}{4\pi i}\ \left(\frac{\beta}{2\pi}\right)^{c/4-8h_{O}}\left(\frac{\bar\beta}{2\pi}\right)^{c/4-8\bar{h}_{O}} e^{\frac{\pi^2 c}{12\beta}+\frac{\pi^2 c}{12\bar\beta}} e^{\frac{\beta}{2}\left(h-\frac{c}{24}\right)+\frac{\bar\beta}{2}\left(\bar h-\frac{c}{24}\right)}
\eeq
which can be evaluated by recognizing the integrals as proportionate to
modified Bessel functions,
\(
I_{\nu}(x) = \frac{1}{2\pi i}\left(\frac{z}{2}\right)^{\nu} \int_{\gamma} \frac{
  du}{u^{\nu+1}} e^{u+\frac{x^2}{4u}}.
\)
This leads to
\beq
K(h,\bar h\rightarrow\infty) \approx
\pi^2 \left(\frac{24h}{c}-1\right)^{\nu/2}\left(\frac{24\bar
    h}{c}-1\right)^{\bar \nu/2} I_{\nu}(x) I_{\bar\nu}(\bar
x)
\eeq
where $\nu = 8h_O-c/4-1$ and
$x=\pi\sqrt{\frac{c}{6}\left(h-\frac{c}{24}\right)}$. Expanding the
Bessel function at large argument,
\beq
K(h,\bar h\rightarrow\infty) \approx  e^{\pi\sqrt{\frac{c}{6}(h-\frac{c}{24})}+ \pi\sqrt{\frac{c}{6}(\bar h-\frac{c}{24})}}
\eeq
and so
\(
\overline{C_{OO(h,\bar h)}^2}|_{h,\bar h\rightarrow \infty} \approx e^{-S_\text{BH}(h)/2+S_\text{BH}(\bar h)/2}
\)
By the logic in section~\ref{generalities}, combined with the two-temperature argument from section 3.1 of \cite{Hartman:2014oaa}, this expression remains valid for all $h,\bar h> c/12$ when $\rho(\Delta)\lesssim e^{\pi \Delta}$.

\subsection*{$\overline{C_{HHL}^2}$: torus two-point function}

Next we generalize the analysis of \cite{Brehm:2018ipf}
to obtain asymptotic results for $\overline{C_{H_1 H_2 L}^2}$ averaged over all operators with dimensions $\Delta_{H_1}$, $\Delta_{H_2}$ and spins $J_{H_1}$, $J_{H_2}$. The logic is the same as in section~\ref{sec:chhl}, but the starting point is the torus two-point function with independent left and right inverse
temperatures $\beta,\bar\beta$ and arbitrary separation between the
operators:

\be
  X(\beta,\bar\beta,z,\bar z)\equiv \tr\left[O(x,t)
                                                    O(0,0) e^{-\beta
                                                    \left(L_0-\frac{c}{24}\right)-\bar\beta\left(\bar{L}_0-\frac{c}{24}\right)}\right]
\ee
where $z=t-x$ and $\bar z=t+x$. In the zero temperature limit,
\be
X(\beta\rightarrow \infty,\bar\beta\rightarrow \infty,z,\bar z)\approx
e^{\frac{\pi c\beta}{12L}} e^{\frac{\pi c\bar\beta}{12L}}\frac{(-1)^{-h_O}\left(\frac{\pi}{L}\right)^{2h_O}}{\sin^{2h_O}\left(\frac{\pi(t-x)}{L}\right)}\frac{(-1)^{-\bar{h}_O}\left(\frac{\pi}{L}\right)^{2\bar{h}_O}}{\sin^{2\bar{h}_O}\left(\frac{\pi(t+x)}{L}\right)}.
\ee
The high-temperature two-point function is
\(
X(\beta\rightarrow 0,\bar\beta\rightarrow 0,z,\bar z)
\approx e^{\frac{\pi c L}{12\beta}} e^{\frac{\pi c L}{12\bar\beta}}
\frac{(-1)^{-h_O}\left(\frac{\pi}{\beta}\right)^{2h_O}}{\sinh^{2h_O}\left(\frac{\pi
      (t-x)}{\beta}\right)} \frac{(-1)^{-\bar{h}_O}\left(\frac{\pi}{\bar\beta}\right)^{2\bar{h}_O}}{\sinh^{2\bar{h}_O}\left(\frac{\pi (t+x)}{\bar\beta}\right)}.
\)
We will take $L=2\pi$. Writing the torus two-point function as a sum over states,
\begin{align}
  \label{2pt2temps}
\langle O(x,t) O(0,0)\rangle_{\beta} &= \sum_{i,j}
                                         |\bra{i}O\ket{j}|^2
                                         e^{i\omega_{ij}(t-x)}e^{i\bar{\omega}_{ij}(t+x)}
                                         e^{-\beta\left(h_i-\frac{c}{24}\right)}
                                         e^{-\bar\beta\left(\bar{h}_i-\frac{c}{24}\right)}\nonumber\\
&= \int_0^{\infty} dh d\bar h \int_{-\infty}^{\infty}d\omega
                                                                                                        d\bar{\omega}\ \ J(h,\bar{h},\omega,\bar{\omega}) e^{i\omega z}e^{i\bar{\omega}\bar{z}} e^{-\beta\left(h-\frac{c}{24}\right)}e^{-\bar{\beta}\left(\bar{h}-\frac{c}{24}\right)}
\end{align}
where $\omega_{ij} = h_i - h_j$, $\bar{\omega}_{ij} = \bar h_i - \bar
h_j$ and the spectral density is
\(
J(h,\bar h,\omega,\bar\omega) = \sum_{i,j} |\bra{i}O\ket{j}|^2
\delta(h_i-h)\delta(\bar{h}_i-\bar{h})
\delta\left((h_i-h_j)-\omega\right) \delta\left((\bar{h}_i-\bar{h}_j)-\bar{\omega}\right).
\)
Inverting \eqref{2pt2temps},
\(
J(h,\bar h,\omega,\bar\omega) = \int_{\gamma\times\bar{\gamma}}
\frac{d\beta}{2\pi i}\frac{d\bar\beta}{2\pi i} \int_{-\infty}^{\infty}
\frac{dz}{2\pi}\frac{d\bar z}{2\pi}e^{-i\omega z}e^{-i\bar{\omega}\bar{z}}
e^{\beta \left(h-\frac{c}{24}\right)} e^{\bar\beta \left(\bar{h}-\frac{c}{24}\right)} X(\beta,\bar\beta,z,\bar z).
\)
As $h,\bar h\rightarrow \infty$ with $\omega,\bar{\omega}$ fixed, the integral will
be dominated by $\beta,\bar{\beta}\rightarrow 0$. Using the high-temperature form of
the thermal two-point function and following the computation of the integrals in
\cite{Brehm:2018ipf} for the scalar case,
\(
  J(h,\bar h,\omega,\bar\omega)|_{h,\bar h\rightarrow \infty}\approx
  I(h,\omega)\cdot I(\bar h,\bar\omega)
  \)
  where
 \begin{align}
  I(h,\omega) &= \sqrt{2}\pi \left(\frac{24}{c}\right)^{h_O-\frac{3}{4}} \frac{\left(h_\text{avg}-\frac{c}{24}\right)^{h_O-\frac 5 4}}{\Gamma(2h_O)} e^{2\pi\sqrt{\frac{c}{6}\left(h_\text{avg} - \frac{c}{24}\right)}}\left|\Gamma\left(h_O+\frac{i\omega}{\sqrt{24 h_\text{avg}/c-1}}\right)\right|^2.
\end{align}
Here $h_\text{avg} = h+\omega/2$ and the exponential factor is the
Cardy density of states for dimensions
$(h_\text{avg},\bar{h}_\text{avg})$. Finally, since
\(
J(h,\bar h,\omega,\bar{\omega}) = \overline{C_{(h,\bar{h})O ((h+\omega,\bar h+\bar\omega))}^2}
\rho(h) \rho(h+\omega) \rho(\bar{h}) \rho(\bar{h}+\bar{\omega}),
\)
we have
\(
\label{asymp-hhl2t}
\overline{C_{(h,\bar{h})O ((h+\omega,\bar h+\bar\omega))}^2}_{h,\bar{h}\rightarrow \infty} \approx \ e^{-S\left(h_\text{avg}\right)-S\left(\bar{h}_\text{avg}\right)}\left|\Gamma\left(h_O+\frac{i\omega}{\sqrt{24h_\text{avg}/c-1}}\right)\right|^2 \left|\Gamma\left(\bar{h}_O+\frac{i\bar{\omega}}{\sqrt{24\bar{h}_\text{avg}/c-1}}\right)\right|^2.
\)
This agrees with the bulk analysis of emission from a spinning BTZ black hole \cite{Maldacena:1997ih}.

If $O$ is light one can apply the logic of section~\ref{sec:chhl} to obtain the two point function at $(\beta,\bar\beta)>2\pi$: under the assumptions of~\cite{Kraus:2017kyl}, the leading form of the two point function is the thermal AdS boundary-to-boundary propagator \cite{Hemming:2002kd}

\be
X(\beta>2\pi,\bar\beta>2\pi,z,\bar z) \approx
e^{\frac{\pi^2 c}{6\beta}} e^{\frac{\pi^2 c}{6\bar\beta}}
\sum_{n=-\infty}^{\infty}\frac{(-1)^{-h_O}\left(\frac{\pi}{\beta}\right)^{2h_O}}{\sinh^{2h_O}\left(\frac{\pi
      (z-2\pi n)}{\beta}\right)}
\frac{(-1)^{-\bar{h}_O}\left(\frac{\pi}{\bar\beta}\right)^{2\bar{h}_O}}{\sinh^{2\bar{h}_O}\left(\frac{\pi
      (\bar z+2\pi n)}{\bar\beta}\right)}
\ee
and so
\begin{align}
  J(h,\bar h,\omega,\bar\omega)|_{h,\bar h > c/12}&\approx
    J(h,\bar h,\omega,\bar\omega)|_{h,\bar h \rightarrow\infty}\cdot \sum_{n=-\infty}^{\infty} e^{2\pi i n(\omega-\bar\omega)}\nonumber\\
&= J(h,\bar h,\omega,\bar\omega)|_{h,\bar h \rightarrow\infty}\cdot \sum_{n=-\infty}^{\infty} \delta\left((\omega-\bar\omega)-n\right).
\end{align}
Eq. \eqref{asymp-hhl2t} therefore continues to hold for all $h,\bar{h}>c/12$ under their assumptions.

\subsection*{$\overline{C_{HHL}}$: torus one-point function}

Finally we generalize the analysis of \cite{Kraus:2016nwo}
to obtain asymptotic results for $\overline{C_{HH L}^2}$ averaged over all operators with dimension $\Delta_{H}$ and spin $J_{H}$. The logic follows section~\ref{sec:spots} except we take independent left and right inverse
temperatures:

\(
\langle O\rangle_{\beta,\bar\beta} = \sum_{\text{states }i}\bra{i}O\ket{i} e^{-\beta\left(h_i-\frac{c}{24}\right) -\bar\beta\left(\bar{h}_i-\frac{c}{24}\right)} =
\int_{0}^{\infty} dhd\bar h\ T(h,\bar h) e^{-\beta\left(h-\frac{c}{24}\right) -\bar\beta\left(\bar{h}-\frac{c}{24}\right)} 
\)
where
\(
T(h,\bar h) = \sum_{\text{states }i} C_{iOi}
\delta(h-h_i) \delta(\bar{h}-\bar{h}_i)=\overline{C_{(h,\bar h) O(h,\bar h)}} \rho(h,\bar h).
\)
The S-transform of $\langle O\rangle_{\beta,\bar\beta}$ is
\(
\langle O\rangle_{\beta,\bar\beta} =  \tau^{-h_O} {\bar\tau}^{-\bar{h}_O}\langle
O\rangle_{\beta',\bar{\beta}'}.
\)
At zero temperature
\(
\langle O\rangle_{\beta,\bar\beta\rightarrow\infty} \approx \bra{\chi}O\ket{\chi}
e^{-\beta \left(h_{\chi}-\frac{c}{24}\right) -\beta \left(\bar{h}_{\chi}-\frac{c}{24}\right)}
\)
where $\chi$ is the operator with smallest $\Delta_{\chi}$ for which
$\bra{\chi}O\ket{\chi}\neq 0$, while at infinite temperature,
\(
\langle O\rangle_{\beta\rightarrow 0} \approx i^s
\bra{\chi}O\ket{\chi}\left(\frac{2\pi}{\beta}\right)^{h_O}\left(\frac{2\pi}{\bar\beta}\right)^{\bar{h}_O}
e^{-\frac{4\pi^2}{\beta}\left(h_{\chi}-\frac{c}{24}\right) -\frac{4\pi^2}{\bar\beta}\left(\bar{h}_{\chi}-\frac{c}{24}\right)}
\)
where $s$ is the spin of $O$. Then
\bea
T(h,\bar h\rightarrow \infty)&\approx&
\int_{\gamma\times\bar\gamma}\frac{d\beta}{2\pi i} \frac{d\bar\beta}{2\pi i}\ i^s
\bra{\chi}O\ket{\chi}\left(\frac{2\pi}{\beta}\right)^{h_O} \left(\frac{2\pi}{\bar\beta}\right)^{\bar{h}_O}
e^{-\frac{4\pi^2}{\beta}\left(h_{\chi}-\frac{c}{24}\right)-\frac{4\pi^2}{\bar\beta}\left(\bar{h}_{\chi}-\frac{c}{24}\right)}e^{\beta \left(h-\frac{c}{24}\right)+\bar\beta \left(\bar{h}-\frac{c}{24}\right)}
\cr
&\approx& \mathcal{N} C_{\chi O \chi}\ e^{2\pi\sqrt{\frac{c}{6}\left(1-\frac{24h_{\chi}}{c}\right)\left(h-\frac{c}{24}\right)}}e^{2\pi\sqrt{\frac{c}{6}\left(1-\frac{24\bar{h}_{\chi}}{c}\right)\left(\bar{h}-\frac{c}{24}\right)}}
 \eea
 where
 \(
 \mathcal{N} = \frac{i^{\bar{h}_O-{h}_O}}{2}
 \frac{(h-c/24)^{h_O-3/4}}{(c/24-h_\chi)^{h_O-1/4}}
 \frac{(\bar{h}-c/24)^{\bar{h}_O-3/4}}{(c/24-\bar{h}_\chi)^{\bar{h}_O-1/4}}.
 \)
This leads to
an asymptotic expression for the average OPE coefficient \cite{Kraus:2016nwo},
\(
\overline{C_{(h,\bar h) O(h,\bar h)}}|_{h,\bar{h}\rightarrow\infty} \approx C_{\chi O \chi}\ e^{-\frac{\pi
    c}{6}\left(1-\sqrt{1-\frac{24
        h_{\chi}}{c}}\right)\sqrt{\frac{24h}{c}-1}}e^{-\frac{\pi
    c}{6}\left(1-\sqrt{1-\frac{24
        \bar{h}_{\chi}}{c}}\right)\sqrt{\frac{24\bar{h}}{c}-1}}.
\)
If we take $c$ large with
the dimensions of $\chi$ fixed,
\(
\overline{C_{(h,\bar h) O(h,\bar h)}}|_{h,\bar{h}\rightarrow\infty} \approx C_{\chi O \chi}\  e^{-2\pi h_{\chi}\sqrt{\frac{24h}{c}-1}} e^{-2\pi \bar{h}_{\chi}\sqrt{\frac{24\bar{h}}{c}-1}}.
\)
This extends to all $h,\bar{h}>c/12$ exactly as in the previous subsection.

\bibliographystyle{toine}
\bibliography{extendedOPE_draft}

\providecommand{\href}[2]{#2}\begingroup\raggedright\begin{thebibliography}{10}

\bibitem{Cardy:1986ie}
J.~L. Cardy, \emph{{Operator Content of Two-Dimensional Conformally Invariant
  Theories}}, Nucl. Phys. {\bf B270} (1986)
186--204

\bibitem{Hartman:2014oaa}
T.~Hartman, C.~A. Keller  and B.~Stoica, \emph{{Universal Spectrum of 2d
  Conformal Field Theory in the Large c Limit}}, JHEP {\bf 09} (2014) 118,
\href{http://www.arXiv.org/abs/1405.5137}{{\tt 1405.5137}}

\bibitem{Das:2017cnv}
D.~Das, S.~Datta  and S.~Pal, \emph{{Universal asymptotics of three-point
  coefficients from elliptic representation of Virasoro blocks}}, Phys. Rev.
  {\bf D98} (2018), no.~10, 101901,
\href{http://www.arXiv.org/abs/1712.01842}{{\tt 1712.01842}}

\bibitem{Brehm:2018ipf}
E.~M. Brehm, D.~Das  and S.~Datta, \emph{{Probing thermality beyond the
  diagonal}}, Phys. Rev. {\bf D98} (2018), no.~12, 126015,
\href{http://www.arXiv.org/abs/1804.07924}{{\tt 1804.07924}}

\bibitem{Cardy:2017qhl}
J.~Cardy, A.~Maloney  and H.~Maxfield, \emph{{A new handle on three-point
  coefficients: OPE asymptotics from genus two modular invariance}}, JHEP {\bf
  10} (2017) 136,
\href{http://www.arXiv.org/abs/1705.05855}{{\tt 1705.05855}}

\bibitem{Kraus:2016nwo}
P.~Kraus and A.~Maloney, \emph{{A cardy formula for three-point coefficients or
  how the black hole got its spots}}, JHEP {\bf 05} (2017) 160,
\href{http://www.arXiv.org/abs/1608.03284}{{\tt 1608.03284}}

\bibitem{Mukhametzhanov:2019pzy}
B.~Mukhametzhanov and A.~Zhiboedov, \emph{{Modular Invariance, Tauberian
  Theorems, and Microcanonical Entropy}},
\href{http://www.arXiv.org/abs/1904.06359}{{\tt 1904.06359}}

\bibitem{Witten:1998zw}
E.~Witten, \emph{{Anti-de Sitter space, thermal phase transition, and
  confinement in gauge theories}}, Adv. Theor. Math. Phys. {\bf 2} (1998)
  505--532, \href{http://www.arXiv.org/abs/hep-th/9803131}{{\tt
  hep-th/9803131}},
[,89(1998)]

\bibitem{Kraus:2017kyl}
P.~Kraus, A.~Sivaramakrishnan  and R.~Snively, \emph{{Black holes from CFT:
  Universality of correlators at large c}}, JHEP {\bf 08} (2017) 084,
\href{http://www.arXiv.org/abs/1706.00771}{{\tt 1706.00771}}

\bibitem{Maldacena:2015iua}
J.~Maldacena, D.~Simmons-Duffin  and A.~Zhiboedov, \emph{{Looking for a bulk
  point}}, JHEP {\bf 01} (2017) 013,
\href{http://www.arXiv.org/abs/1509.03612}{{\tt 1509.03612}}

\bibitem{Haehl:2014yla}
F.~M. Haehl and M.~Rangamani, \emph{{Permutation orbifolds and holography}},
  JHEP {\bf 03} (2015) 163,
\href{http://www.arXiv.org/abs/1412.2759}{{\tt 1412.2759}}

\bibitem{Belin:2014fna}
A.~Belin, C.~A. Keller  and A.~Maloney, \emph{{String Universality for
  Permutation Orbifolds}}, Phys. Rev. {\bf D91} (2015), no.~10, 106005,
\href{http://www.arXiv.org/abs/1412.7159}{{\tt 1412.7159}}

\bibitem{Belin:2015hwa}
A.~Belin, C.~A. Keller  and A.~Maloney, \emph{{Permutation Orbifolds in the
  large N Limit}}, Annales Henri Poincare (2016) 1--29,
\href{http://www.arXiv.org/abs/1509.01256}{{\tt 1509.01256}}

\bibitem{Maldacena:1997ih}
J.~M. Maldacena and A.~Strominger, \emph{{Universal low-energy dynamics for
  rotating black holes}}, Phys. Rev. {\bf D56} (1997) 4975--4983,
\href{http://www.arXiv.org/abs/hep-th/9702015}{{\tt hep-th/9702015}}

\bibitem{Hemming:2002kd}
S.~Hemming, E.~Keski-Vakkuri  and P.~Kraus, \emph{{Strings in the extended BTZ
  space-time}}, JHEP {\bf 10} (2002) 006,
\href{http://www.arXiv.org/abs/hep-th/0208003}{{\tt hep-th/0208003}}

\bibitem{Pal:2019yhz}
S.~Pal, \emph{{Bound on asymptotics of magnitude of three point coefficients in
  2D CFT}},
\href{http://www.arXiv.org/abs/1906.11223}{{\tt 1906.11223}}

\bibitem{Maxfield:2019hdt}
H.~Maxfield, \emph{{Quantum corrections to the BTZ black hole extremality bound
  from the conformal bootstrap}},
\href{http://www.arXiv.org/abs/1906.04416}{{\tt 1906.04416}}

\bibitem{Calabrese:2009ez}
P.~Calabrese, J.~Cardy  and E.~Tonni, \emph{{Entanglement entropy of two
  disjoint intervals in conformal field theory}}, J. Stat. Mech. {\bf 0911}
  (2009) P11001,
\href{http://www.arXiv.org/abs/0905.2069}{{\tt 0905.2069}}

\bibitem{Calabrese:2010he}
P.~Calabrese, J.~Cardy  and E.~Tonni, \emph{{Entanglement entropy of two
  disjoint intervals in conformal field theory II}}, J. Stat. Mech. {\bf 1101}
  (2011) P01021,
\href{http://www.arXiv.org/abs/1011.5482}{{\tt 1011.5482}}

\bibitem{Cho:2017oxl}
M.~Cho, S.~Collier  and X.~Yin, \emph{{Recursive Representations of Arbitrary
  Virasoro Conformal Blocks}}, JHEP {\bf 04} (2019) 018,
\href{http://www.arXiv.org/abs/1703.09805}{{\tt 1703.09805}}

\bibitem{Cho:2017fzo}
M.~Cho, S.~Collier  and X.~Yin, \emph{{Genus Two Modular Bootstrap}}, JHEP {\bf
  04} (2019) 022,
\href{http://www.arXiv.org/abs/1705.05865}{{\tt 1705.05865}}

\bibitem{Mathur:1988xc}
S.~D. Mathur and A.~Sen, \emph{{Differential Equation for Genus Two Characters
  in Arbitrary Rational Conformal Field Theories}}, Phys. Lett. {\bf B218}
  (1989)
176--184

\bibitem{Witten:2007kt}
E.~Witten, \emph{{Three-Dimensional Gravity Revisited}},
\href{http://www.arXiv.org/abs/0706.3359}{{\tt 0706.3359}}

\bibitem{Gaiotto:2007xh}
D.~Gaiotto and X.~Yin, \emph{{Genus two partition functions of extremal
  conformal field theories}}, JHEP {\bf 08} (2007) 029,
\href{http://www.arXiv.org/abs/0707.3437}{{\tt 0707.3437}}

\bibitem{Belin:2017nze}
A.~Belin, C.~A. Keller  and I.~G. Zadeh, \emph{{Genus two partition functions
  and Rényi entropies of large c conformal field theories}}, J. Phys. {\bf
  A50} (2017), no.~43, 435401,
\href{http://www.arXiv.org/abs/1704.08250}{{\tt 1704.08250}}

\bibitem{Zamolodchikov:1985ie}
A.~B. Zamolodchikov, \emph{{CONFORMAL SYMMETRY IN TWO-DIMENSIONS: AN EXPLICIT
  RECURRENCE FORMULA FOR THE CONFORMAL PARTIAL WAVE AMPLITUDE}}, Commun. Math.
  Phys. {\bf 96} (1984)
419--422

\bibitem{Zamolodchikov1987}
A.~B. Zamolodchikov, \emph{Conformal symmetry in two-dimensional space:
  Recursion representation of conformal block}, Theoretical and Mathematical
  Physics {\bf 73} (Oct, 1987)
1088--1093

\bibitem{shouvik_note}
S.~Datta. {Private communication}.

\bibitem{Asplund:2014coa}
C.~T. Asplund, A.~Bernamonti, F.~Galli  and T.~Hartman, \emph{{Holographic
  Entanglement Entropy from 2d CFT: Heavy States and Local Quenches}}, JHEP
  {\bf 02} (2015) 171,
\href{http://www.arXiv.org/abs/1410.1392}{{\tt 1410.1392}}

\bibitem{Shenker:2013pqa}
S.~H. Shenker and D.~Stanford, \emph{{Black holes and the butterfly effect}},
  JHEP {\bf 03} (2014) 067,
\href{http://www.arXiv.org/abs/1306.0622}{{\tt 1306.0622}}

\bibitem{Anous:2019yku}
T.~Anous and J.~Sonner, \emph{{Phases of scrambling in eigenstates}}, SciPost
  Phys. {\bf 7} (2019) 003,
\href{http://www.arXiv.org/abs/1903.03143}{{\tt 1903.03143}}

\bibitem{Fitzpatrick:2014vua}
A.~L. Fitzpatrick, J.~Kaplan  and M.~T. Walters, \emph{{Universality of
  Long-Distance AdS Physics from the CFT Bootstrap}}, JHEP {\bf 08} (2014) 145,
\href{http://www.arXiv.org/abs/1403.6829}{{\tt 1403.6829}}

\bibitem{Fitzpatrick:2015zha}
A.~L. Fitzpatrick, J.~Kaplan  and M.~T. Walters, \emph{{Virasoro Conformal
  Blocks and Thermality from Classical Background Fields}}, JHEP {\bf 11}
  (2015) 200,
\href{http://www.arXiv.org/abs/1501.05315}{{\tt 1501.05315}}

\bibitem{Harlow:2011ny}
D.~Harlow, J.~Maltz  and E.~Witten, \emph{{Analytic Continuation of Liouville
  Theory}}, JHEP {\bf 12} (2011) 071,
\href{http://www.arXiv.org/abs/1108.4417}{{\tt 1108.4417}}

\bibitem{Pappadopulo:2012jk}
D.~Pappadopulo, S.~Rychkov, J.~Espin  and R.~Rattazzi, \emph{{OPE Convergence
  in Conformal Field Theory}}, Phys. Rev. {\bf D86} (2012) 105043,
\href{http://www.arXiv.org/abs/1208.6449}{{\tt 1208.6449}}

\end{thebibliography}\endgroup

\end{document}